# Modeling of a vertical tunneling graphene heterojunction field-effect transistor

S. Bala Kumar, Gyungseon Seol and Jing Guo[a)]

Department of Electrical and Computer Engineering, University of Florida, Gainesville, FL, 32611

[*]Corresponding author. E-mail: a) guoj@ufl.edu

**Abstract**

Vertical tunneling field-effect-transistor (FET) based on graphene heterojunctions with layers of $h$BN is simulated by self-consistent quantum transport simulations. It is found that the asymmetric p-type and n-type conduction is due to work function deference between the graphene contact and the tunneling channel material. Modulation of the bottom-graphene-contact plays an important role in determining the switching characteristic of the device. Due to the electrostatic short-channel-effects stemming from the vertical-FET structure, the output I-V characteristics do not saturate. The scaling behaviors the vertical-FET as a function of the gate insulator thickness and the thickness of the tunneling channel material are examined.

Graphene is an interesting material with unusual properties[1]. This makes graphene a promising material for future electronic devices.[2] One of the proposed applications of graphene is the graphene-based field-effect-transistor (FET). However, the absence of bandgap in graphene would result in a high leakage current, and thus it would be difficult to turn off the transistor[3]. This effect is further worsen by Klein tunneling[4] which allow electron tunneling across a potential barrier. As a result the ON-OFF ratio in a graphene based transistor is highly limited. Many methods have been proposed to overcome this problem, i.e. pattering of graphene nanoribbon[5], applying electric field across multilayer graphene[6], using quantum dots[7] or chemical derivates[8]. Recently an alternative vertical-tunneling-graphene (VTG) FET design was demonstrated[9]. This device, in which graphenes are used as contacts and not the channel, functions based on quantum tunneling across a thin insulating channel barrier such as layers of $h$BN or $MoS_2$, sandwiched between a pair of graphene contacts.

In this paper, we theoretically model the VTG-FET, investigate the qualitative behavior of the experimental observations and suggest methods for device optimization. We use a self-consistent capacitance model to obtain the electrochemical potential profile across the device, and then using non-equilibrium-greens-function (NEGF)[10] method, we compute the current and conductance across the device. Unlike conventional transistors, where the source-drain contact simply play a passive role as reservoir of electrons, the operation of this device is primarily determined by the modulation of density-of-state (DOS) in the contacts, which is tuned by the applied gate and source-drain voltage. Due to the short-channel-effect, arising from the vertical structure, the I-V output of this device does not saturate. We also investigate the effect of channel and insulator thickness on optimizing performance of this device.

Our modeled device consists of a tunneling channel made of layered-hBN sandwiched between two graphene layers as shown in Fig. 1(a), similar to the structure demonstrated by L. Britnell et al.[9] The graphene layers act as the source and drain contacts. The source-drain bias, $V_b$ and the bottom gate voltage, $V_g$ are applied as in Fig. 1(a). First we use a capacitance model to calculate the electrochemical potential at graphene and the hBN layers. Here, we explain the computation for a device with the channel consisting of a monolayer hBN. The charge at each layer-i, $Q_i$ is related to the vacuum energy level, $E_i$ as follows:

$$-Q_i = C_i \left( E_{i-1}^{vac} - E_i^{vac} \right) + C_{i+1} \left( E_{i+1}^{vac} - E_i^{vac} \right), \quad i = 0,1,2,3,4 \tag{1}$$

,where the capacitance $C = \varepsilon_0 \varepsilon_r / d$, $\varepsilon_r$ is the dielectric constant and $d$ is the thickness. Referring to Fig 1(b) and setting the boundaries $C_0 = C_4 = 0$ and $E_{-1}^{vac} = E_4^{vac} = 0$, Eq. 1 can be expressed as

$$\begin{bmatrix} C_1 & -C_1 & & \\ -C_1 & C_1+C_2 & -C_2 & \\ & -C_2 & C_2+C_3 & -C_3 \\ & & -C_3 & C_3 \end{bmatrix} \begin{bmatrix} E_0^{vac} \\ E_1^{vac} \\ E_2^{vac} \\ E_3^{vac} \end{bmatrix} = \begin{bmatrix} -eQ_0 \\ -eQ_1 \\ -eQ_2 \\ -eQ_3 \end{bmatrix}, \text{ and further simplified to}$$

$$\begin{bmatrix} E_1^{vac} \\ E_2^{vac} \\ E_3^{vac} \end{bmatrix} = \begin{bmatrix} C_1+C_2 & -C_2 & \\ -C_2 & C_2+C_3 & -C_3 \\ & -C_3 & C_3 \end{bmatrix}^{-1} \begin{bmatrix} -eQ_1 + C_1 E_0^{vac} \\ -eQ_2 \\ -eQ_3 \end{bmatrix} \tag{2}$$

Note that $E_0^{vac} = -eV_g + \phi_g$, where $\phi_g$ is the bottom gate work function. From the value of $E_i^{vac}$ obtained in Eq. 2, the charge density, $Q_i$ is computed

$$Q_i = -e \int_{U_i}^{\infty} D_i(E) f_i^e(E) dE + e \int_{-\infty}^{U_i} D_i(E) f_i^p(E) dE \tag{3}$$

$U_i = E_i^{vac} - \phi_i$ is the source Dirac-point energy ($E_S^{dir}$), channel mid-gap energy ($E^{mid}$), and drain Dirac-point energy ($E_D^{dir}$) when i=1,2, and 3 respectively. D is the density-of-state (DOS) and $f^{e(p)}$ is the electron (hole) occupancy. Eq. 2 and 3 is then solved self-consistently until $U_i$ converges. For hBN with more number of layers, the above equations are modified accordingly. The $U_i$ profile across the device at flat band condition (when an arbitrary Vg and Vb is applied) is illustrated in Fig. 1(c) (Fig. 1(d)). Unless otherwise stated we use the following parameters for the simulation: SiO$_2$ insulator thickness, $t_{ins}$=300nm and SiO$_2$ dielectric $\varepsilon_{ins}$=4. The work function of the Si gate, graphene source/drain contacts and hBN channel are $\phi_g$=4eV, $\phi_{SD}$=4.7eV[11] and $\phi_C$=3eV, respectively. The distance between graphene contacts and the hBN channel is 0.5nm. hBN interlayer distance is 0.35nm[12]. Number of channel layers, N=3. Note that $U_0$=-eV$_g$, $U_1 = E_S^{dir}$, and $U_{N+3} = E_D^{dir}$. $U_2$ to $U_{N+2}$ is equal to the $E^{mid}$ of each hBN layers.

The electron transport behavior across the device is studied by using NEGF formalism[10]. The Hamiltonian of the monolayer hBN using π-orbital tight bonding model is given by

$$H(K) = \begin{bmatrix} E^{mid} + E_{gap}/2 & -t_0 - 2t_0 e^{+iK_x} \cos K_y \\ -t_0 - 2t_0 e^{-iK_x} \cos K_y & E^{mid} - E_{gap}/2 \end{bmatrix} \quad (4)$$

where $K_y = \sqrt{3} k_y a_0 / 2$, $K_x = 3 k_x a_0 / 2$, intralayer nearest-neighbor (NN) hoping parameter, $t_0$=2.3eV[12], bandgap of the monolayer hBN, $E_{gap}$=5eV[13], and the NN intralayer atomic distance $a_0$=0.15nm. For multilayer hBN, we use AB-stacking hBN layers with interlayer hoping parameter, $t_p$=0.6eV[12]. In the NEGF formalism, the electron transport across the device for a given energy E is

$$T(E) = \sum_K Tr[\Gamma_S(E)G(E,K)\Gamma_D(E)G(E,K)^\dagger] \tag{5}$$

where $G(E,K) = [EI - H(K) - \Sigma_S(E) - \Sigma_D(E)]^{-1}$ is the retarded Green's function of the GNR channel, $\Sigma(E)$ is the self energy coupling to the graphene contacts, and $\Gamma(E) = i[\Sigma(E) - \Sigma^+(E)]$. The subscript S(D) refers to the source (drain) contacts. For the self energy, the top and bottom, hBN layers are connected to the respective graphene contacts with a contact broadening of $\sigma = \beta|E - E^{dir}|L(E^{dir}, \gamma)$, where $L(E^{dir}, \gamma) = \frac{\gamma/\pi}{(E - E^{dir})^2 + \gamma^2}$ is the Lorentzian broadening function. Unless otherwise stated, we use $\beta = 0.01$ and $\gamma = 0.01 eV$. Note that this expression is obtained by using the linear DOS of graphene, where the DOS is zero at the Dirac point, $E^{dir}$. Therefore the transmission is minimum when $E = E^{dir}$. We define $T_S(E) = T(E)/S$, S is the total area of the Brillion zone used in Eq. 5. At a finite temperature, $\tau$ the areal-conductance across the graphene, at Fermi energy, $E_F$ is

$$G_D(E_F) = g_0 \int_{-\infty}^{+\infty} T_S(E_F) \frac{1}{4k_B\tau} \text{sech}^2\left(\frac{E - E_F}{2k_B\tau}\right) dE \tag{6}$$

, where $g_0 = e^2/h$ and $k_B$ is the Boltzmann constant. When a finite bias is applied, the areal-source-drain current,

$$I_{SD} = g_0 \int_0^{qV_b} T_S(E)(f_S(E) - f_D(E)) dE \tag{7}$$

,where Fermi distribution function, $f_{S/D}(E) = \left(1 + \exp\left(\frac{E - \mu_{S/D}}{k_b\tau}\right)\right)^{-1}$, $\mu_S = 0$, and $\mu_D = eV_b$. The applied $V_b$ and $V_g$ changes the H(K) in Eq. 4, and thus varies $G_D$ and $I_{SD}$.

Fig. 2(a) shows the $G_D$-$V_g$ curve obtained from our calculation. These results show a very good qualitative agreement with the experimental results from ref. 9, which is replotted in Fig. 2(b). We first investigate the ambipolar behavior observed in the $G_D$-$V_g$ curve. One interesting feature of this device is that the qualitative behavior of the transport is strongly determined by the graphene bottom-contact, unlike in conventional transistors where the contacts play only a passive role. Therefore, in this VTG-FET, contact modulation is more critical, compared to channel barrier modulation. The non-uniform DOS in graphene, in contrast to the constant DOS in metallic contacts, results in the contact modulation. In graphene, DOS is zero at the Dirac point, and hence minimum transmission is obtained at the $E^{dir}$, even though in intrinsic $h$BN layers, the minimum transmission is obtained at the $E^{mid}$. Thus, as shown in Fig. 2(c-e), the alignment of graphene Dirac points relative to the $E_F$ strongly affects the electron transport across the heterojunction between the bottom graphene contact and the channel.

Since Fig. 2(d) shows the zero bias conductivity, $G_D$. i.e. $V_b=0$, the Fermi energy is located at $E_F=\mu_S=\mu_D=0$. Referring to Fig. 2 (c-e), the minimum conductance, which is proportional to the minimum leakage current, is obtained when the Fermi level, $E_F$ is aligned with the graphene Dirac point, i.e. $E_S^{dir} = E_D^{dir} = 0$. This condition is achieved when $V_g$=$V_{FB}$≈-1eV, where $V_{FB}$ is the flat band (FB) voltage. When |$V_g$-$V_{FB}$|>0, the DOS in graphene contacts become larger, and thus the $G_D$ increases. Therefore we see an ambipolar behavior. We further study the effect of the quality of the contacts. Quantitatively, the quality of the contacts is model by increasing the broadening of the DOS in the contacts, $\gamma$. As shown in the inset of Fig. 2 (c), when the $\gamma$ increases the ambipolar behavior gradually disappears, while the minimum leakage current increases. For example, at $\gamma$ =0.1eV the $G_D$ decreases monotonically with increasing $V_g$, and the minimum leakage current is 10 times larger than when $\gamma =0.01eV$. The contact modulation, as well as the minimum leakage current is also enhanced by the quality of the contact. In the experimental results in ref. 9, $h$BN is used as the insulator in between $SiO_2$ and graphene. It has been shown that graphene grown on a $h$BN substrate are of high quality[14]. Furthermore, graphene also forms a good contact with $h$BN channel due to similarity in atomic structure. Therefore, due to this high quality contact, the modulation of the contact DOS strongly affects the conductance across the $h$BN.

Another interesting feature in the Fig. 2(a), is the asymmetry of the $G_D$-$V_g$ curve. The magnitude of the gradient of the $G_D$-$V_g$ curve is not the same for $V_g<V_{FB}$ and $V_g>V_{FB}$. This asymmetric behavior is mainly caused by the different work function of the graphene contact and the channel material. As shown in Fig. 3(a-c), at FB, if $\phi_{SD} > \phi_C$ ($\phi_{SD} < \phi_C$), the Fermi level is closer to the valence (conductance) band, and hence the channel layer is a p(n)-type semiconductor. As a result, higher conductance is obtained when Vg<$V_{FB}$(Vg>$V_{FB}$). In our structure, $\phi_{SD} > \phi_C$, and thus the Fermi level is closer to the valence band. Therefore at FB, the channel is a p-type semiconductor, in which the transmission is mainly due to the holes and thus gradient is higher when $V_g<V_{FB}$. We further quantify the asymmetry of the curve as $\alpha = -g_m^- / g_m^+ -1$, where $g_m^-$ ($g_m^+$) is the gradient of $G_D$-$V_g$ curve when $V_g<V_{FB}$ ($V_g>V_{FB}$). Note that $\alpha>0$ ($\alpha<0$) indicates the curve is asymmetric with larger gradient when $V_g<V_{FB}$ ($V_g>V_{FB}$) and $\alpha=0$ indicates that the curve is perfectly symmetric. Referring to Fig. 3(d-e), when we hypothetically vary the work function of the channel, the asymmetry is varied. When $\phi_{SD} = \phi_C$, where the channel is an instrinsic semiconductor at the FB, the curve is symmetric, i.e. $\alpha=0$. On the hand, when $\phi_{SD} < \phi_C$ ($\phi_{SD} > \phi_C$), the channel becomes a n-type semiconductor at FB, and thus $\alpha < 0$ ($\alpha > 0$) indicating the conductance is higher for when $V_g > V_{FB}$ ($V_g < V_{FB}$). Note that in all these cases we only considered the case where at the FB the Fermi level is within the channel band gap. We also would like to highlight that a similar effect of work function difference on $G_D$-$V_g$ asymmetric has also been observed in horizontal nanoribbon FET, when the different contact materials are used[15].

Fig. 4(a) shows the $I_{SD}$-$V_b$ curve calculated from our model, while Fig. 4(b) shows the $I_{SD}$-$V_b$ curve from the experimental results[9]. A good qualitative agreement is observed. One striking feature of these results is that, unlike conventional FETs, the $I_{SD}$ does not saturate. In horizontal, electrostatically well designed FETs, the gate bias has much stronger control over the channel compared to the drain bias, which leads to a saturation of source-drain current. However, short channel length and stronger drain control degrades the saturation output characteristics. This phenomenon is known as drain-induced-barrier-lowering (DIBL) which occurs due to short-channel-effect in conventional FETs. A similar phenomenon is observed in this VTG-FET. Here, due to the ultra-thin channel length, i.e. in the order of 1nm, and due to the structure of the device, the gate voltage has less control over the channel potential. Therefore the drain bias causes significant change in the channel barrier, and thus the $I_{SD}$ does not saturate. An effect reminiscent of DIBL is shown in Fig. 4(c-e), where the channel barrier is lowered by the increasing $V_b$, resulting in increasing current. It is also worth nothing that the electronic transport in this VTG-FET is primarily due to the tunneling current. Referring to Fig. 4(c-e), as we increase the $V_b$ the transmission window remains within the tunneling regime. The increasing of current with increasing $V_b$ is caused by the decrease in the tunneling height.

Finally, we investigate the effect of channel thickness and bottom-gate insulator thickness, $t_{ins}$. The effect of channel thickness, which is proportional to the number of $h$BN layers N, is shown in Fig. 5(a). The increase of N is equivalent to an increase in barrier width and thus the electron tunneling decreases, exponentially decreasing the $G_D$. On the other hand, the increase in $t_{ins}$, does not change the minimal leakage current [Fig. 5(b)]. This is because, at FB, where the minimal leakage current is obtained, the contact potential is align with the Fermi level, $E_F=0$. The variation in $t_{ins}$ does not effect this alignment, and thus the potential barrier within the device is independent of the $t_{ins}$. However, when the $V_g \neq V_{FB}$, the $t_{ins}$ determines the potential profile across the device. The gate controbality of this potential profile increases with decreasing $t_{ins}$, resulting in a higher $G_D$, and thus increasing the ON-OFF ratio.

In conclusion, we study the electronic transport in a VTG-FET. This FET exhibits an ambipolar behavior similar to the recent experimental results. We also found an asymmetric p-type and n-type conduction is due to work function deference between the graphene contact and the tunneling channel material. The thin channel layer in VTG-FET, leads to a short channel effect, and thus the $I_{SD}$ do not saturate when $V_b$ is increased. Finally we showed that the increase in $t_{ins}$ decreases the minimum leakage current, while the variation in insulator thickness does not affect the minimum leakage current. The modeling work not only explains the major features observed in a recent experiment, but also indicates the importance of improving the quality of the bottom graphene contact and scaling down the gate insulator of the VTG-FET.

We would like to thank Prof. P. Kim of Columbia University for technical discussions. This work was supported by NSF and ONR.

**Figure and Caption**

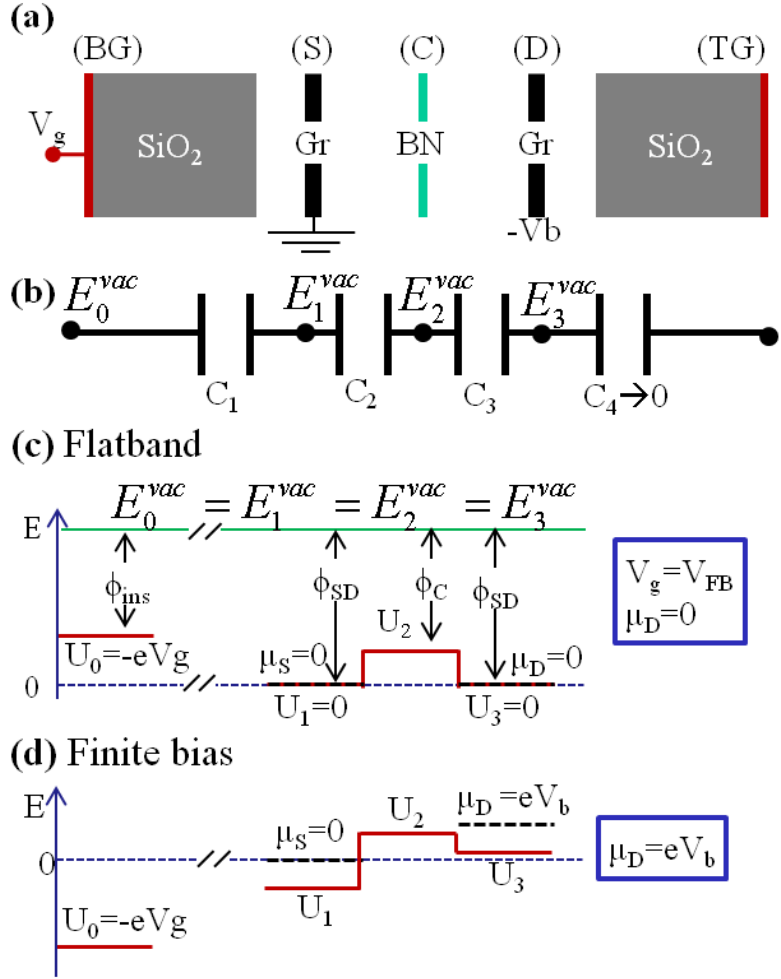

Figure 1 (a) Shows the device structure with a monolayer $h$BN as the channel. TG, S, C, D, and BG represent top-gate source, channel, drain, and bottom-gate respectively. Gr refers to graphene. (b) The capacitance model corresponding to (a). $E^{vac}$ is the vacuum energy level. The potential profile across the device (c) at flatband condition, and (d) when an arbitrary Vg and Vb is applied. $\phi_g$, $\phi_{SD}$, $\phi_C$ refers to the gate, source/drain, and channel workfunctions, respectively. The BG potential, $U_0=-eV_g$, Fermi energy at source (drain), $\mu_S=0$ ($\mu_D=eVb$). The potential definitions are consistent with ref. 9. $U_{1(3)}$ is the graphene Dirac point energy at source (drain), while $U_2$ is the midgap energy of the channel.

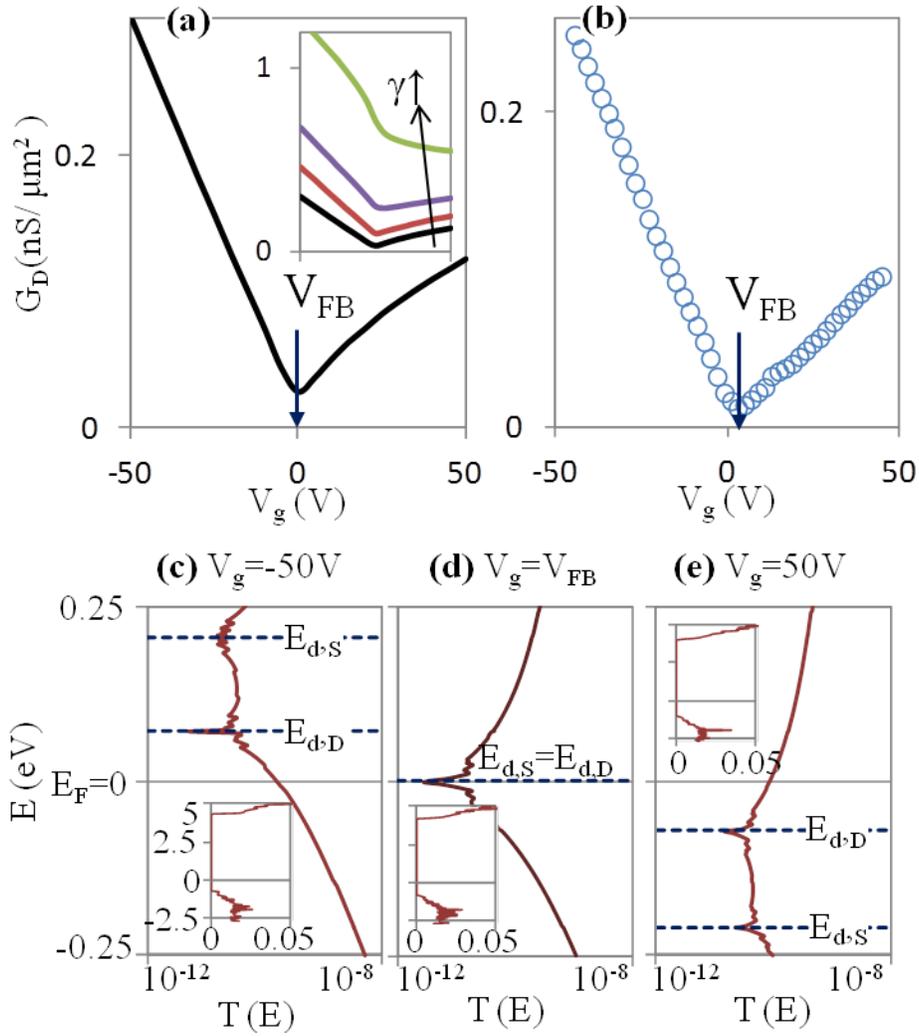

Figure 2 (a) $G_d$-$V_g$ curve at Vb=0. The inset shows $G_d$ with varying contact broadening, γ [γ=0.01, 0.05, 0.1, 0.5eV]. (b) Experimental results obtain in ref. 9, showing a good qualitative agreement with (a). The transmission curve in log scale for (c) $V_g$<$V_{FB}$, (d) $V_g$=$V_{FB}$, and (e) $V_g$>$V_{FB}$ at the vicinity of Fermi level, $E_F$=0. The inset in (c-e) show the transmission in linear scale for a larger energy range.

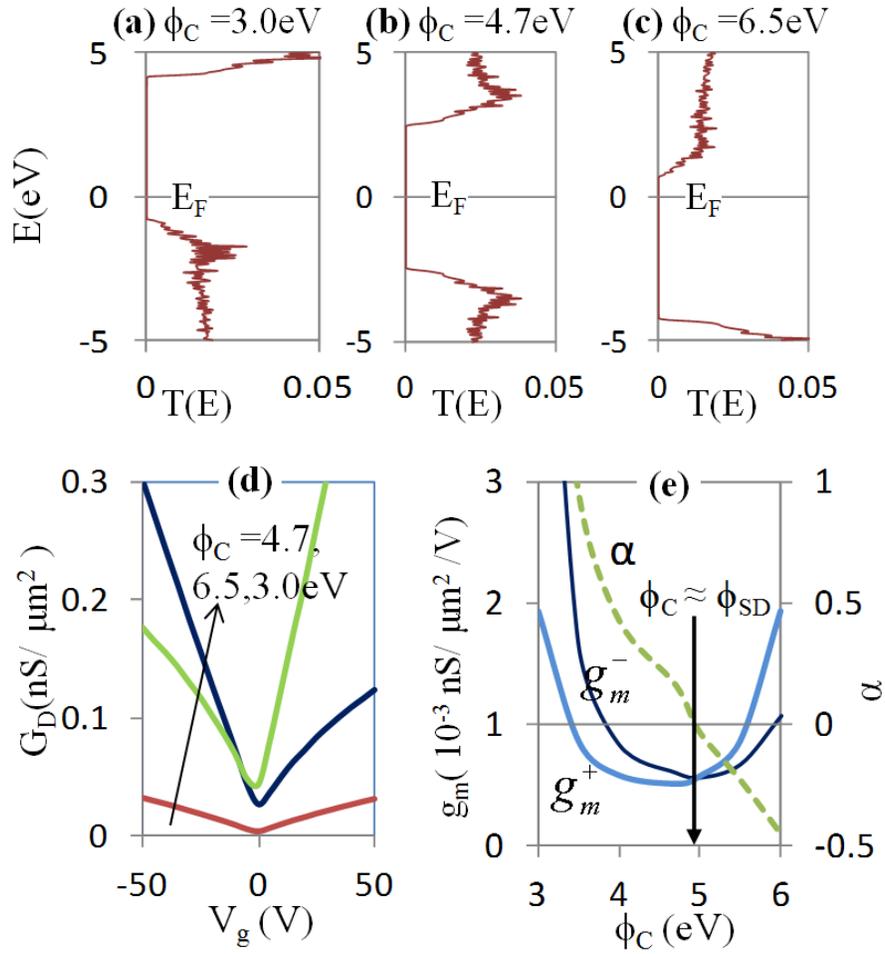

Figure 3 Transmission curve at FB for (a) $\phi_C < \phi_{SD}$, (b) $\phi_C = \phi_{SD}$, (c) $\phi_C > \phi_{SD}$. The channel varies from a p-type to n-type semiconductor as the $\phi_C$ increases. (d) Shows $G_D$-$V_g$ for different $\phi_C$ values. (e) $g_m^+$, $g_m^-$, and α variation as a function of $\phi_C$. When $\phi_C \approx \phi_{SD}$, $g_m^-=g_m^+$ resulting in α =0, i.e. a symmetric $G_D$-$V_g$ curve.

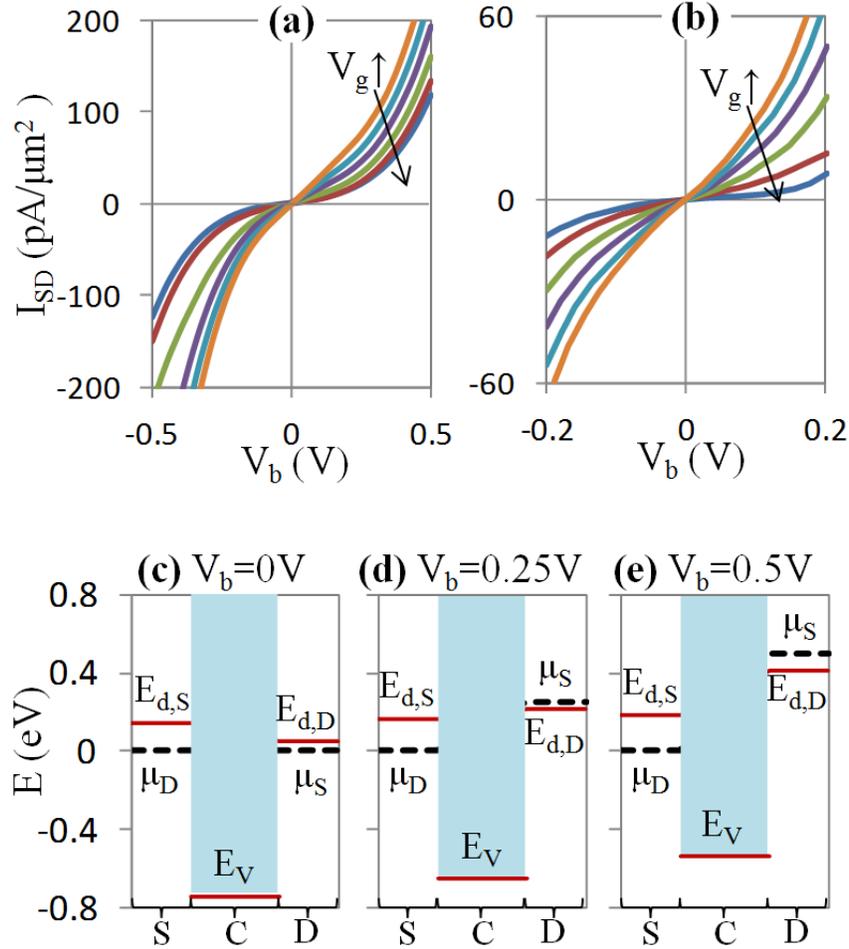

Figure 4 (a) $I_{SD}$-$V_b$ curve for different $V_g$ values. $V_g$ is varied from -45V to 5V in a step of 10V. (b) Experimental results obtain in ref. 9, showing a good qualitative agreement with (a). The potential profile across the device when (c) $V_b$=0V, (d) $V_b$=0.25V, and (e) $V_b$=0.5V. The shaded region indicates band gap. The channel potential is decreases with increasing $V_b$, reminiscent of drain-induced-barrier-lowering effect (DIBL).

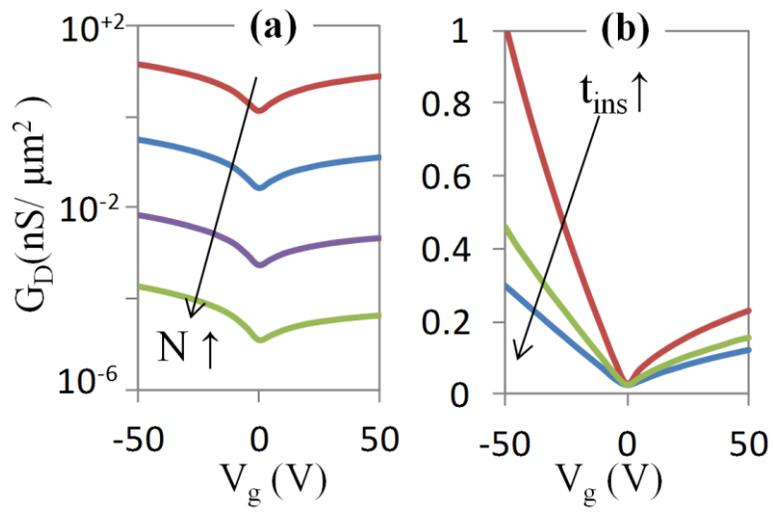

Figure 5 (a) Shows $G_D$-$V_g$ for different N values. [N=2, 3, 4, 5]. (b) Shows $G_D$-$V_g$ for different $t_{ins}$ values. [$t_{ins}$= 100, 200, 300 nm].